\documentstyle[11pt]{article}
\def\sun{\hbox{$\odot$}}
\def\acknowledgements{\vspace{12pt}\noindent{\em Acknowledgements.\/
}%
\ignorespaces}

\begin{document}
\setcounter{page}{0}
\title{\bf The Edge of the Galactic Disc}
\thanks{Based on observations made at
Canada-France-Hawaii Telescope (CFHT)}
\author{\small Annie C. Robin$^{1,2}$, Michel Cr\'ez\'e$^{2}$,
Vijay Mohan$^{3}$\\
{\small $^{1}$Observatoire de Besan\c{c}on, BP1615, F-25010
Besan\c{c}on
Cedex, France} \\
{\small  $^{2}$CNRS URA1280, Observatoire de Strasbourg,
11 rue de l'Universit\'e,}\\
{\small F-68000 Strasbourg, France} \\
{\small $^{3}$U.P. State Observatory, Manora Peak, Nainital, 263129 India
}}

\maketitle

\begin{abstract}
As part of a stellar population sampling program, a series of
photometric probes at various field sizes and depths have been obtained in a
low extinction window in the galactic anticentre direction. Such data set
strong constraints on the radial structure of the disc. At the forefront of
this "drilling" program, very deep CCD frames probe the most
external parts of the disc. Over the whole effective magnitude range (18 to
25),
all contributions in the statistics which should be expected from old disc
stars
beyond 6 kpc vanish, although such stars dominate by far at distances less than
5 kpc. This is the signature of a sharp cut-off in the star density: the edge
of the galactic disc between 5.5 and 6 kpc. As a consequence, the galactic
radius does not exceed 14 kpc (assuming $R_{\sun}$=8.5). Colours of
elliptical
galaxies measured in the field rule out the risk of being misled by undetected
extinction.
\end{abstract}
\vskip 1cm
{\bf Subject headings:} Galaxy: structure - Galaxy:
stellar content\\\\

\section{Introduction}
Efforts to sample star distributions in the galactic plane should face
both the problem of overcrowding and the complexe structure of the
absorbing layer. For these reasons the outer part of our own galactic disc is
very poorly known. The radial scale length of the density decrease is
controversial and we have nearly no indication what happens
at the end.

Two previous papers (Mohan et al. 1988, and Robin at al. 1992)
present the first results of a stellar population sampling program,
including a series of photometric probes at various field sizes and
depths in a low extinction window in the galactic anticentre direction.
Wide field photometry in UBV in the
magnitude range 12-17 is shown to interpret unambiguously in terms of
extinction and stellar density. The interpretation partially uses the
model developped by Robin and Cr\'ez\'e (1986) and Bienaym\'e et al.
(1987). The model ingredients which play a role in the present
anticentre investigation are the density law of the galactic disc
(radially exponential) and the luminosity function from Wielen et al.
(1983). Strong constraints are set on the radial
structure of the disc: the galactic disc scale length is 2.5 kpc.
Based on this scale length and assuming no dramatic change in the
luminosity function one can predict what should normally happen at
faintest magnitudes: the  V, B-V distribution of stars resulting from
this investigation is given in figure 1a and the V counts are plotted in
figure 2 (dashed line).

\section{Deep observations towards the anticentre}

Deep CCD observations in the BV bands have been obtained at the 3.6
meter CFH Telescope in a low extinction window at low latitude in the
direction of the galactic anticentre.
A detailed description of observational and data analysis aspects is
given elsewhere (Robin et al. 1992).
Observations cover four
neighbouring fields adding up to 29 square arcminutes. Field
characteristics are given in table 1. The detection limit
is about magnitude 28 in V, while the completeness limit corresponding
to a photometric accuracy better than 0.1 is 25 in V and 22.5 to 24 in B
depending on the frame. The resulting V, B-V distribution is plotted in
figure 1b down to V = 23. B data are lacking beyond this limit.
Differential star counts down to V = 25 are plotted on figure 2
together with the associated 1 sigma error bars.

\section{Interpretation}

The faint star count predictions deviate strongly from the observations at the
faint end and there is a clear excess of blue stars in the down left part of
figure 1a. The bulk of disc contributors in this magnitude range is made of
disc
dwarfs beyond 5.5 kpc (in figure 1a stars closer than 5.5 kpc are plotted as
full circles while farther stars are crosses).

All disagreements in the V, B-V diagram vanish if the stellar galactic
disc ends abruptly at 5.5 kpc as also shown in the V star
counts in figure 2 (solid line). Other explanations could come from the
extinction along the line of sight, the scale length used in the
model or the luminosity function. We show below that none of these
hypotheses resists to investigation.

Interpretation of the cutoff seen in the V, B-V diagram in terms of
absorbing cloud would need a total extinction A$_V$ = 2.4 over the line
of sight, including
a cloud of A$_V$ = 1.2 at 5.5 kpc that
is about 300 pc from the galactic plane. The extinction measured from
the UBV diagrams and HI column density gives a total extinction to 4 kpc
of A$_{V}$ = 1.2. The colours of 4 elliptical galaxies measured in the
fields give an estimate of the total extinction inside the Galaxy of
1.4, incompatible with the existence of a deeply absorbing cloud a 5.5
kpc. This is also in agreement with the fact that the fields are inside
Special Area 23 selected by Kapteyn as a low extinction window.

Density distributions with scale lengths larger than the adopted 2.5
kpc would impose a still closer cutoff (at 3.5 kpc from us if h = 3.5 kpc)
while
shorter scale lengths are hardly compatible with the observations of bright
stars in the same region (Mohan et al. 1988).

\section{Discussion}

Stars that should be expected to appear in the sample to V=23 at distances
larger than 5.5 kpc are common dwarfs of absolute visual magnitudes between 3
and 7, a part of the luminosity function which is well known from the study of
the solar neighbourhood. Moreover these stars appear at brighter V magnitudes
when they are closer. Changing their luminosity function would lead to strong
disagreements with Schmidt plates data of Mohan et al. at magnitudes 12 to 16
and with the CCD data at magnitudes 17 to 21. Rather than invoking a sudden
change of the luminosity function of stars of magnitude 3 to 7 at 5.5 kpc from
us the data are better interpreted as a sharp cutoff of the density
distribution
for all stars at this distance. This observation is indicative of the
end of the star formation process, possibly related to the external Lindblad
resonnance.

Most external disc galaxies show a radial truncation. These
cutoffs seem to arise within 1 kpc or less (van der Kruit 1988). From a
sample of 7 edge-on galaxies this drop has been found between 3.4 and 5.3
times the old disc scale length (van der Kruit and Searle 1981, 1982). When
reanalysing the Wevers' sample (1986) of 20 face-on spirals, van der Kruit
(1988)
showed that the mean ratio between $R_{max}$ and scale length h
is 4.5 $\pm$ 1.0 for 16 galaxies and larger than 6.0 for the remaining 4
spirals. However Barteldrees and Dettmar (1990) sample give a mean ratio
of 2.8 $\pm$ 1. It should be noted that the scale lengths of
external spirals may be uncertain by a factor two (Knappen and van der Kruit
1991).

In our Galaxy star counts in the anticentre (Robin et al. 1992) give a
radial scale length of 2.5 $\pm$ 0.3 kpc. Together with the presently
determined cutoff it implies a ratio $R_{max} / h$ of 5.6 $\pm$ 0.6 if
$R_{\sun}$ = 8.5 kpc or 5.2 $\pm$ 0.6 if $R_{\sun}$ = 7.5 kpc, in
agreement with the observed ratio in external galaxies.

Indications for a cutoff in the disc distribution has already been found
in other data. From the IRAS Point Source Catalogue Habing (1988) found a
cutoff at distances between 1 and 2.5 kpc. However Habing stressed that
this very short cut-off distance depends strongly on the assumed disc
scale length (4.5 kpc), which on the other hand is inconsistent with our
wide field star count results. Habing did not test any model
with a shorter scale length for the old disc components. In the view of
the present result the combination of short scale and larger distance
cutoff should be tested against IRAS data.

The edge of the disc was also determined in the gaseous component of the Galaxy
at larger distances. Wouterloot et al. (1990) got a sharp decline of the CO
cloud
density between 18 and 20 kpc, a signature that star formation stops at about
this distance. However these distances being derived from kinematics are
subject
to errors due to perturbations of the velocity field. HII regions observed by
Fich et al. (1989) also end at galactocentric distances of about 15 kpc.

Our determination of the radial extent of the old disc does not conflict
with the possibly larger extent of young stars or star forming regions
if an evolutionnary scenario like the one of Larson (1976) (where the
star formation propagates from the centre of the Galaxy to the outer
part) is realistic. In this case one expects to find only recent star
formation in the outer part of the Galaxy and no old disc stars.\\\\

\acknowledgements{This research was partially supported by the Indo-French
Centre for the Promotion of Advanced Research / Centre Franco-Indien
Pour la Promotion de la Recherche Avanc\'ee.}\\\\\\

{\noindent \large \bf References}\\\\
Barteldrees, A., \& Dettmar, R. J. 1990, in {\it International Conference on
Dynamics
and Interactions of galaxies}, ed R. Wielen (Springer), p. 348\\
Bienaym\'e, O., Robin, A. C. \& Cr\'ez\'e, M. 1987, {\it Astron.
Astrophys.}
{\bf 180}, 94\\
Fich, M., Blitz, L. \& Stark, A. A. 1989, {\it Astrophys. J.} {\bf 342},
272\\
Habing, H. J. 1988, {\it Astron. Astrophys.} {\bf 200},40\\
Knappen, J. H. \& van der Kruit, P. C. 1991, {\it Astron. Astrophys.}
{\bf 248}, 57\\
Larson, R. B. 1976, {\it Mon. Not. R. astr. Soc.} {\bf 176}, 31\\
Mohan, V., Bijaoui, A., Cr\'ez\'e, M. \& Robin, A. C. 1988, {\it Astron.
Astrophys. Suppl.} {\bf 73}, 85\\
Robin, A. C. \& Cr\'ez\'e, M. 1986, {\it Astron. Astrophys.} {\bf
157}, 71\\
Robin, A. C., Cr\'ez\'e, M. \& Mohan, V. 1992, {\it Astron. Astrophys.}
in press\\
van der Kruit, P. C. 1988, {\it Astron. Astrophys.} {\bf 192}, 117 \\
van der Kruit, P. C. \& Searle, L. 1981, {\it Astron. Astrophys.} {\bf
95}, 105\\
van der Kruit, P. C. \& Searle, L. 1982, {\it Astron. Astrophys.} {\bf
110}, 61\\
Wevers, B. M. H. R., van der Kruit, P. C., \& Allen, R.J. 1986, {\it Astron.
Astrophys. Suppl.} {\bf 66}, 505\\
Wielen, R. Jahreiss, H. \& Kr\"uger, R. 1983, In {\it The Nearby Stars and
the Stellar Luminosity Function}, I.A.U Coll. 76, p. 163 , Davis Philip and
Upgren (eds), L. Davis Press (Schenectady, NY).\\
Wouterloot, J. G. A., Brand, J., Burton, W. B. \& Kwee, K. K. 1990, {\it
Astron. Astrophys.} {\bf 230}, 21\\\\\\

\newpage
{\noindent \bf Table 1:} Field coordinates and area in square arc minutes.
\begin{flushleft}
\begin{tabular}{|l|lllll|}
\hline
Field  &  RA (1950) & Dec (1950) & l & b & Area\\
\hline
1       &5 53 06.5&+30 39 43&   179.70 & 2.88 & 6.7\\
2       &5 52 45.4&+30 40 52&   179.64 & 2.83 & 6.5\\
3       &5 52 54.7&+30 36 13&   179.73 & 2.82 & 7.7\\
4       &5 52 46.4&+30 38 24&   179.68 & 2.81 & 6.6\\
\hline
\end{tabular}
\end{flushleft}
\vskip 3cm
{\noindent \large \bf Figure captions}\\\\

\noindent {\bf Figure 1:} (V, B-V) diagram of anticentre stars. (a)
Model predicted distribution. Dots are stars closer than 5.5 kpc while
crosses are stars beyond 5.5 kpc. (b) Observed distribution. The solid
line is a guide to identify the zone where most stars are beyond 5.5
kpc. \\

\noindent {\bf Figure 2:} V star counts to magnitude 25 in the
anticentre direction. Diamonds: Observed counts with 1 sigma error bars
(Poisson noise only). Dashed line: Predicted counts assuming no cutoff
in the radial distribution of stars. Solid line: Predicted counts
assuming a cutoff at 5.5 kpc from the sun. \\

\newpage
{\noindent \large \bf Authors postal addresses}\\\\

{\bf \noindent Michel Cr\'ez\'e}, Observatoire de Strasbourg, 11 rue de
l'Universit\'e, F-68000 Strasbourg, France \\
{\bf \noindent Vijay Mohan}, U.P. State Observatory, Manora Peak, Nainital,
263129
India\\
{\bf \noindent Annie C. Robin}, Observatoire de Besan\c{c}on, BP1615,
F-25010
Besan\c{c}on Cedex, France\\

\end{document}